\documentclass[sigconf]{acmart}

\usepackage{wrapfig}
\usepackage{multirow}
\usepackage{longtable}
\usepackage{xspace}
\usepackage{tabularray}
\usepackage{tabularx}
\usepackage{enumitem}
\usepackage{caption}
\usepackage{subcaption}
\usepackage{graphicx}
\usepackage{placeins} 

\AtBeginDocument{%
  }

\usepackage{tabularx}      
\usepackage{array}         
\newcolumntype{Y}{>{\raggedright\arraybackslash}X} 

\copyrightyear{2025}
\acmYear{2025}
\acmDOI{}

\acmConference[GenAICHI 2025]{Generative AI and HCI: A CHI 2025 Workshop}{April 27, 2025}{Yokohama, Japan and Online}
\acmISBN{}

\usepackage{rotating}
\usepackage{tabularx}
\usepackage{graphicx}  
\usepackage{tikz}      
\usepackage{xcolor} 

\begin{document}

\title[LACE]{LACE: Exploring Turn-Taking and Parallel Interaction Modes in Human-AI Co-Creation for Iterative Image Generation}

\author{Yen-Kai Huang}
\affiliation{%
  \institution{Dartmouth College}
  \city{Hanover}
  \state{New Hampshire}
  \country{USA}
}
\email{yenkai.huang.gr@dartmouth.edu}

\author{Zheng Ning}
\affiliation{%
  \institution{University of Notre Dame}
  \city{Notre Dame}
  \state{Indiana}
  \country{USA}
}
\email{zning@nd.edu}

\author{Ming Cheng}
\affiliation{%
  \institution{Dartmouth College}
  \city{Hanover}
  \state{New Hampshire}
  \country{USA}}
\email{ming.cheng.gr@dartmouth.edu}



\begin{abstract}
Generative AI (GenAI) has transformed visual art creation by allowing rapid ideation and prototyping from text prompts. However, it falls short in meeting professional artists’ needs due to their limited control over specific image elements, lack of coherence in iterative refinement, and incompatibility with established workflows. To address these challenges, we present LACE, a Human-AI co-creative system for iterative image generation with professional creative tools. Inspired from the Co-Creative Framework for Interaction (COFI)~\cite{COFI}, LACE supports both turn-taking and parallel interaction modes for iterative image generation. Through a within-subjects study with 21 participants across three creative tasks, we show preliminary results on how participation modes influence user experience, creative outcomes, and workflow preferences.

\end{abstract}



\begin{teaserfigure}
  \includegraphics[width=\textwidth]{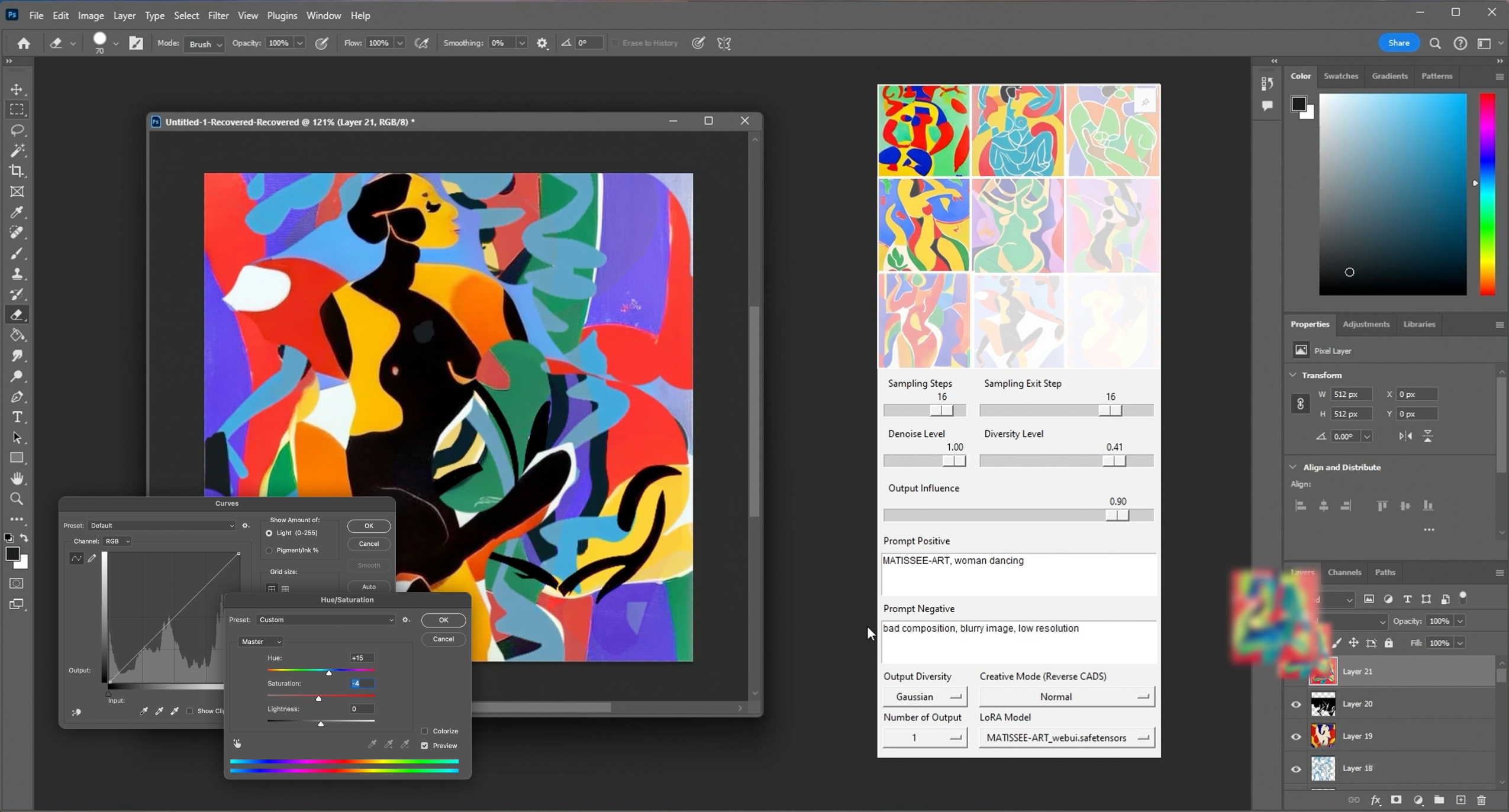}
  \caption{Main interface of LACE, integrated into Adobe Photoshop}
  \Description{Main interface of LACE integrated into Adobe Photoshop, showing image prompting controls, layer management panels, and iterative refinement tools designed for professional artists.}
  \label{fig:teaser}
\end{teaserfigure}

\maketitle

\section{Introduction}
Generative AI systems, such as MidJourney, FLUX, or Stable Diffusion, have surged in popularity for image creation. However, most of these systems still rely on single‐turn workflows. Addressing this limitation, the Co-Creative Framework for Interaction (COFI) identifies participation style—parallel versus turn-taking—as critical for effective human-AI collaboration \cite{COFI}. Although many creative support tools (CSTs) favor parallel interaction, where human and AI contributions occur simultaneously, a review of 92 CST systems revealed 89.1\% adopt parallel interaction, with only 10.9\% incorporating turn-taking \cite{Rezwana2019}. Research suggests that turn-taking can foster deeper mutual understanding through clear role delineation \cite{IkegamiIizuka2007}, whereas parallel interaction often supports greater real-time creative fluency \cite{Fan2019Collabdraw}. Despite these findings, systems that support both interaction modes remain unexplored \cite{Berkel_HAI}. Enabling flexible transitions between parallel and turn-taking behaviors in human–AI co-creation may offer enhanced flexibility and better alignment with professional creative practices.

\begin{figure}
    \centering
    \includegraphics[width=1.\linewidth]{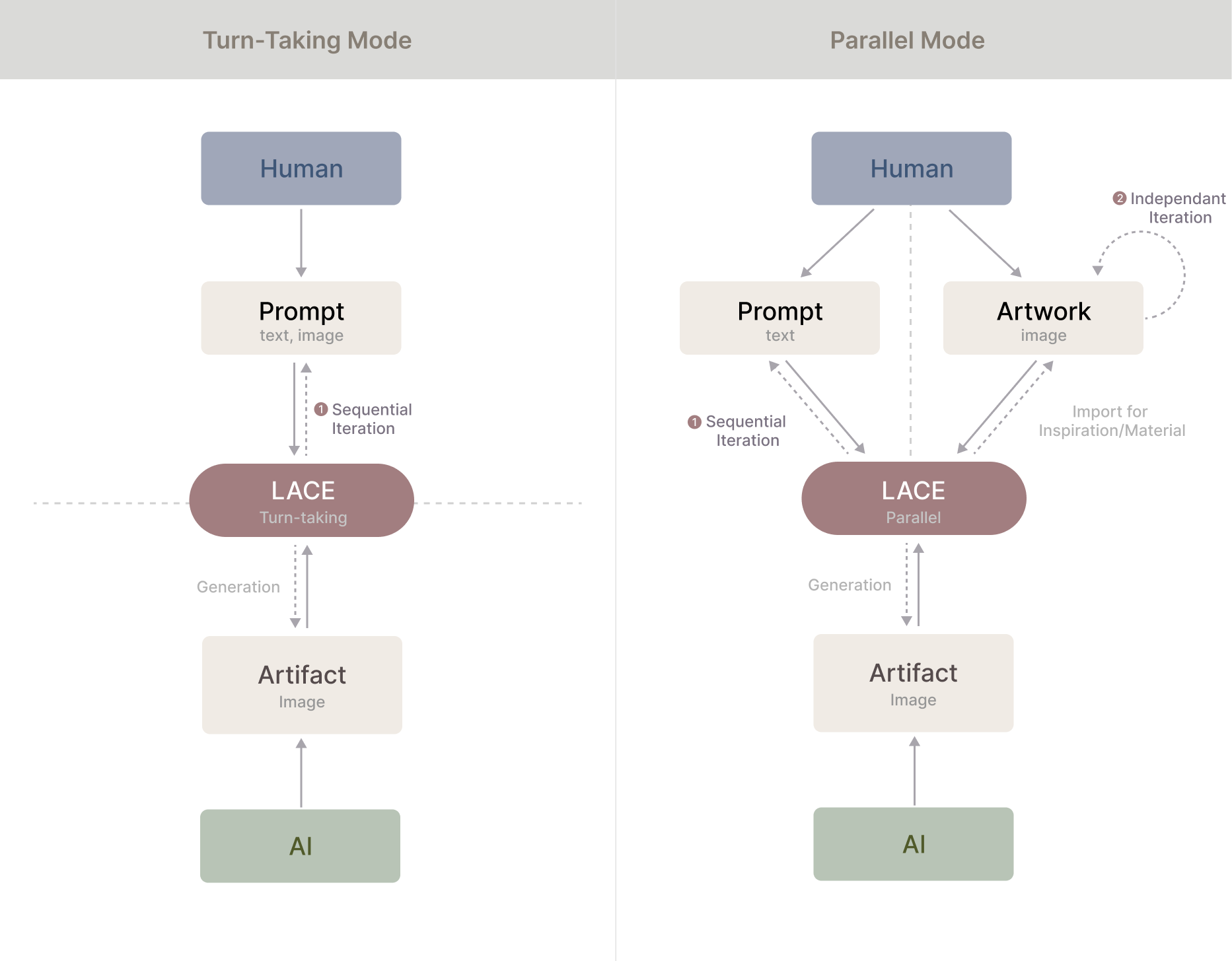}
    \caption{The LACE system supports two interaction modes: Turn-Taking and Parallel. Turn-Taking employs a classic sequential participation style, where human and AI iteratively refine the prompt and generated artifact in sequence. In Parallel mode, LACE maintains the AI loop for prompt iteration while introducing an independent Artist loop, allowing artists to refine their work in real-time without interruption.}
    \label{fig:LACE-mode}
\end{figure}

To address this research gap, we introduce LACE, a system integrated into Adobe Photoshop that supports both parallel and turn-taking human-AI interactions (see figure~\ref{fig:LACE-mode}). In parallel mode, a dual-feedback loop continuously generates AI-driven images, imported as independent layers. This allows artists to refine their work in real-time while receiving new suggestions without overwriting previous edits. In contrast, the turn-taking mode follows a more traditional text-to-image or image-to-image process, where users adjust prompts and receive a single output. Our pilot evaluation (N=21) provides preliminary evidence that such flexible interaction modes can enhance creative control and authorship.

Although this study does not provide a full comparative analysis of parallel versus turn-taking modalities, our findings strongly demonstrate the potential of LACE to support flexible interaction modes, offering valuable insights into how these approaches can enhance creative control and authorship. This work provides guidance for future research to further explore and evaluate the distinct contributions of participation styles in human–AI co-creation.

Our contributions are threefold: 1) we present LACE, a flexible co-creative system that integrates iterative AI pipelines within professional image editing workflows; 2) we offer initial observations on how blending parallel and turn-taking interactions may enhance user engagement and control; and 3) we outline a framework and future research agenda for more rigorous comparisons of participation styles in co-creative AI systems.

\section{RELATED WORK}
    \subsection{Support Creativity Tools}
    Generative AI systems such as MidJourney, FLUX, and Stable Diffusion have sparked considerable interest by producing visuals from text prompts, image prompts or maksing filling with latent diffusion models~\cite{StableDiffusion,MidJourney}. However, most retain a single-turn approach—users provide a prompt, receive an image, and iterate only through prompt tweaking. While effective for rapid prototyping, this style lacks iterative refinement and local control over elements, potentially limiting creative coherence~\cite{gholami2024streamliningimageeditinglayered, lugmayr2022repaint}.

    \subsection{Parallel vs. Turn-Taking in Human-AI collaboration}
    In Human–AI co-creativity research, participation style is widely recognized as critical to effective collaboration~\cite{COFI, Shneiderman_2002, Fischer2004}. The Co-Creative Framework for Interaction (COFI) specifically highlights parallel (where human and AI act simultaneously) and turn-taking (structured, sequential input) as key modes~\cite{COFI}. A large‐scale review of 92 creativity support tools (CST) shows an overwhelming preference for parallel interactions (89.1\%), with only 10.9\% incorporating turn-taking~\cite{Rezwana2019}. Despite indications that structured turn-taking fosters mutual understanding\cite{IkegamiIizuka2007} and that parallel interactions benefit real-time fluency\cite{Fan2019Collabdraw}, comparatively few studies explore the flexibility of blending both styles~\cite{Berkel_HAI}.

    \subsection{Iterative Image Generation}
    Professional tools (e.g., Adobe Photoshop) increasingly integrate AI functionalities, yet they often maintain single‐step or prompt-driven workflows without robust iterative control~\cite{batley2024GFsynergy}. Although advanced features like neural filters and generative fill aid creativity, they rarely allow users to choose between parallel or turn-taking modes fluidly. This gap is especially noteworthy for professional creators seeking to balance spontaneity (parallel collaboration) with deliberation (turn-taking). Our approach, LACE, aims to address these shortcomings by merging continuous AI-driven suggestions with layer-based editing to preserve authorship, providing initial evidence of the potential benefits of flexible participation styles.

\section{LACE SYSTEM}
    \begin{figure}
        \centering
        \includegraphics[width=0.85\linewidth]{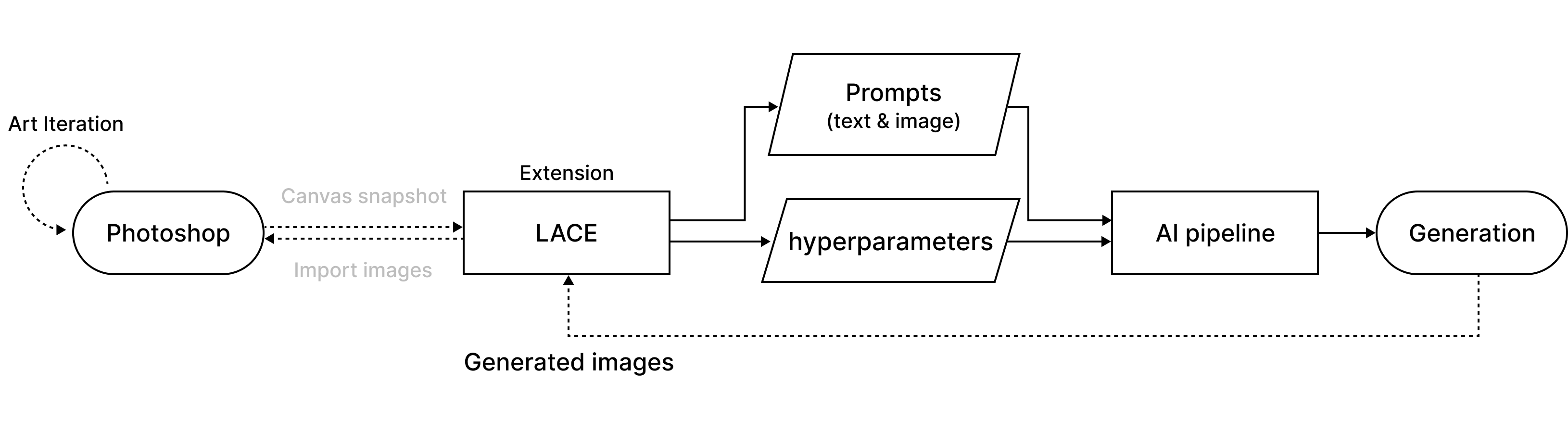}
        \caption{System architecture of LACE}
        \label{fig:LACE-arc}
    \end{figure}
    
    LACE builds on literature in digital art~\cite{okada2024process, daniel2022creative}, AI-assisted image creation~\cite{semantic_segmentation_diffusion, cao2023animediffusion}, and control in generative models~\cite{feng2024item, yang2024fine}. It enables precise control over image generation, supports iterative refinement for creative continuity, and integrates seamlessly into professional workflows by extending Adobe Photoshop.

    Drawing on the Co-Creative Framework for Interaction (COFI) \cite{COFI}, LACE supports two distinct participation styles—turn-taking and parallel—giving creators the flexibility to choose which style best aligns with their current creative task. By unifying these modes in the same Photoshop-based pipeline, LACE leverages industry-standard features such as layers, blending modes, and masking while providing different ways for the AI and user to alternate or synchronize their contributions. Figure~\ref{fig:LACE-mode} illustrates these two modes at a high level and the architecture in figure~\ref{fig:LACE-arc}.
    \begin{enumerate}
        \item 
        \textbf{Turn-taking Collaboration}: Turn-taking mode enables structured workflows where users and the AI alternate contributions sequentially. This mode allows artists to iteratively refine ideas, focusing on precision and coherence. For example, users can provide the model with initial sketches or prompts, receive AI-generated outputs, and build upon these results step by step. This sequential process is particularly suited for tasks requiring strong foundational elements, such as defining compositions or applying major stylistic transformations.
        \item 
        \textbf{Parallel Collaboration}:Parallel interaction mode is designed for real-time exploration and late-stage refinement. This mode allows the AI to continuously generate suggestions in the background while the artist independently refines their work. Artists can selectively import AI-generated outputs as layers, maintaining creative momentum without interruptions. This mode supports fluid experimentation, enabling users to integrate AI-generated elements into their artwork dynamically. Tasks that require detailed adjustments, such as fine-tuning textures or colors, may benefit from this synchronous workflow.
    \end{enumerate}
    
    Turn-taking supports structured refinement for late-stage tasks requiring precision, while parallel interaction enables fluid exploration and rapid ideation, offering dynamic adaptability to creative workflows.

    \subsection{Supporting Both Modes}
    Turn-taking supports structured refinement for late-stage tasks requiring precision, while parallel interaction enables fluid exploration and rapid ideation, offering dynamic adaptability to creative workflows.
    LACE acts as an intermediary layer between Photoshop and AI models, managing the interaction between artist input and AI generation. It captures snapshots of the artist’s canvas and adjusts the influence weight, determining whether the generated images are entirely AI-driven (influence weight of 0) or reflect the artist’s work (influence weight of 1). Generated outputs are cached in LACE, allowing artists to decide which results to import into Photoshop as new layers. Imported outputs then feed back into the AI pipeline, creating a cyclic loop where successive generations build on prior work.

    The system’s GUI does not explicitly switch modes. If results are not imported into Photoshop, LACE operates as a traditional text-to-image tool. When imported, the results integrate into the workflow as part of a layered, iterative refinement process. Turn-taking supports structured refinement by generating outputs step-by-step, ideal for late-stage tasks like style transfer or precision adjustments. Parallel interaction generates real-time suggestions during active work, supporting early-stage ideation and exploration.
    
    LACE enables flexible human-AI collaboration by bridging creative workflows and AI tools, adapting dynamically to both exploratory and refinement-focused tasks while maintaining user control.

    \begin{figure}
        \centering
        \includegraphics[width=1\linewidth]{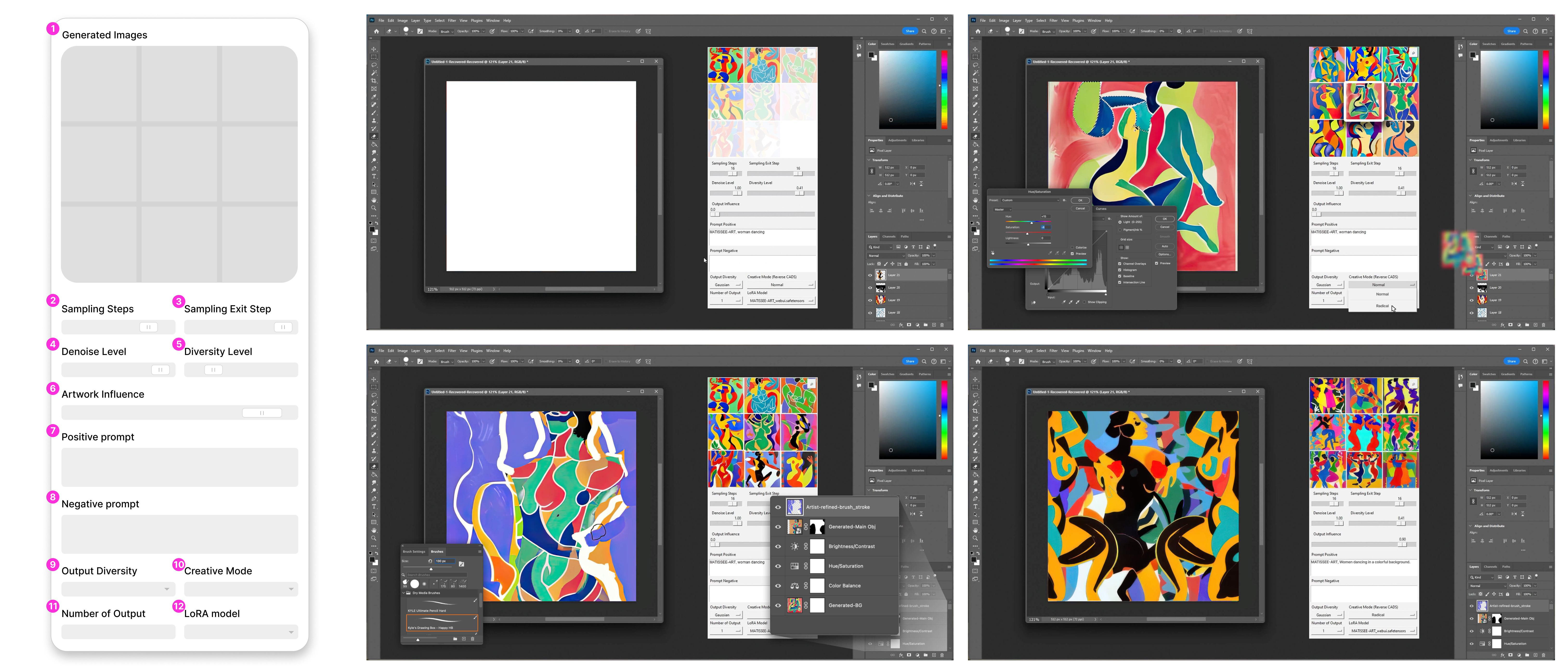}
        \caption{Interface and Workflow of LACE}
        \label{fig:LACE-compilation}
    \end{figure}

\section{Evaluation}
    We conducted a within-subjects experiment to compare two interaction modes in LACE—\textbf{parallel} and \textbf{turn-taking}—across creative image-generation tasks. Each participant produced artwork using both modes, allowing direct comparisons of perceived ownership, usability, satisfaction, expectation, explainability and art score. Qualitative feedback was gathered through open-ended interviews.
    
    For polit test, we recruited twenty-one volunteers (age 20–27, 4 male and 17 female) were recruited via university flyers and social media posts. Their Photoshop experience ranged from less than one year (45\%) to over four years (30\%), with the remainder at an intermediate level. Most had a background in computer science or digital art, and 71.4\% reported prior experience with generative AI tools. Participants received \$25 USD for their time, and the study followed an institutionally approved research protocol.
    
    Each participant was randomly assigned one of the three prompt-driven art tasks: representational, abstract, and a design challenge (see in table~\ref{tab:testing_material}). Participants need to try out all three workflows (see table~\ref{tab:workflow_comparison}) for up to 15 minutes, and participants could stop earlier if they felt satisfied with their results.
    
    After testing each workflow, participants completed a 7-point Likert scale questionnaire assessing ownership, satisfaction, perceived usability, alignment with expectations, and whether they considered the result “art.” They also reported their estimated completion time. A short, semi-structured interview followed each session to capture qualitative insights, preferences, and any challenges encountered.
    
    Participants received a brief tutorial on LACE’s interface and the two interaction modes. For each of the three art prompts, they alternated between the \emph{parallel} mode—where LACE continuously generated new suggestions while they edited—and the \emph{turn-taking} mode—where the AI produced an image output only after participants refined their prompt or images. Mode order was randomized to counterbalance potential learning or fatigue effects. In each mode, they worked until they were satisfied or reached the 15-minute time limit, after which they completed the questionnaire. Finally, participants discussed their overall experiences in a concluding interview.
    
    Quantitative data (ownership, satisfaction, usability, etc.) were analyzed using non-parametric tests (Friedman for overall differences and Wilcoxon signed-rank for pairwise comparisons), with Kendall’s W and Cohen’s reported as effect sizes. Qualitative feedback from interviews underwent a reflexive thematic analysis, enabling identification of recurring patterns and user perceptions of each interaction mode.
    
    \begin{table}[ht]
      \footnotesize
      \label{tab:workflow_comparison}
      \begin{tabularx}{\columnwidth}{|Y|Y|Y|Y|}
        \hline
        \textbf{Workflow} &
        \textbf{Dependency on Previous Result} &
        \textbf{Interaction Style} &
        \textbf{Sampling Process} \\ \hline
        W1: Basic Turn‑Taking &
        No dependency; generates new results each time &
        Sequential turn‑taking &
        Single sampling \\ \hline
        W2: Iterative Turn‑Taking &
        Depends on previous results &
        Sequential turn‑taking &
        Multiple rounds of sampling \\ \hline
        W3: Parallel/Hybrid (LACE) &
        Depends on previous results &
        Parallel with optional turn‑taking &
        Multiple rounds of sampling \\ \hline
      \end{tabularx}
      \caption{Comparison of workflows tested in a within‑subjects design}
    \end{table}
    \label{subsec:workflows}

\begin{table}
    \centering
    \footnotesize  
    \begin{tabularx}{\columnwidth}{|c|Y|}
    \hline
    \textbf{Type} & \textbf{Art Prompt} \\
    \hline
    \textbf{T1: Representational} & A man reaching for a painting in an art gallery, accompanied by a dog sniffing another artwork on the floor. \\
    \hline
    \textbf{T2: Non-Representational} & An abstract composition that embodies the dynamics and motion associated with joy. \\
    \hline
    \textbf{T3: Design Challenge} & A pixel art game scene with a bustling cityscape featuring assorted architectural styles. \\
    \hline
    \end{tabularx}
    \caption{Summary of types and prompts for different tasks}
    \label{tab:testing_material}
\end{table}

\section{RESULT}
    \begin{figure}
        \centering
        \includegraphics[width=1.\linewidth]{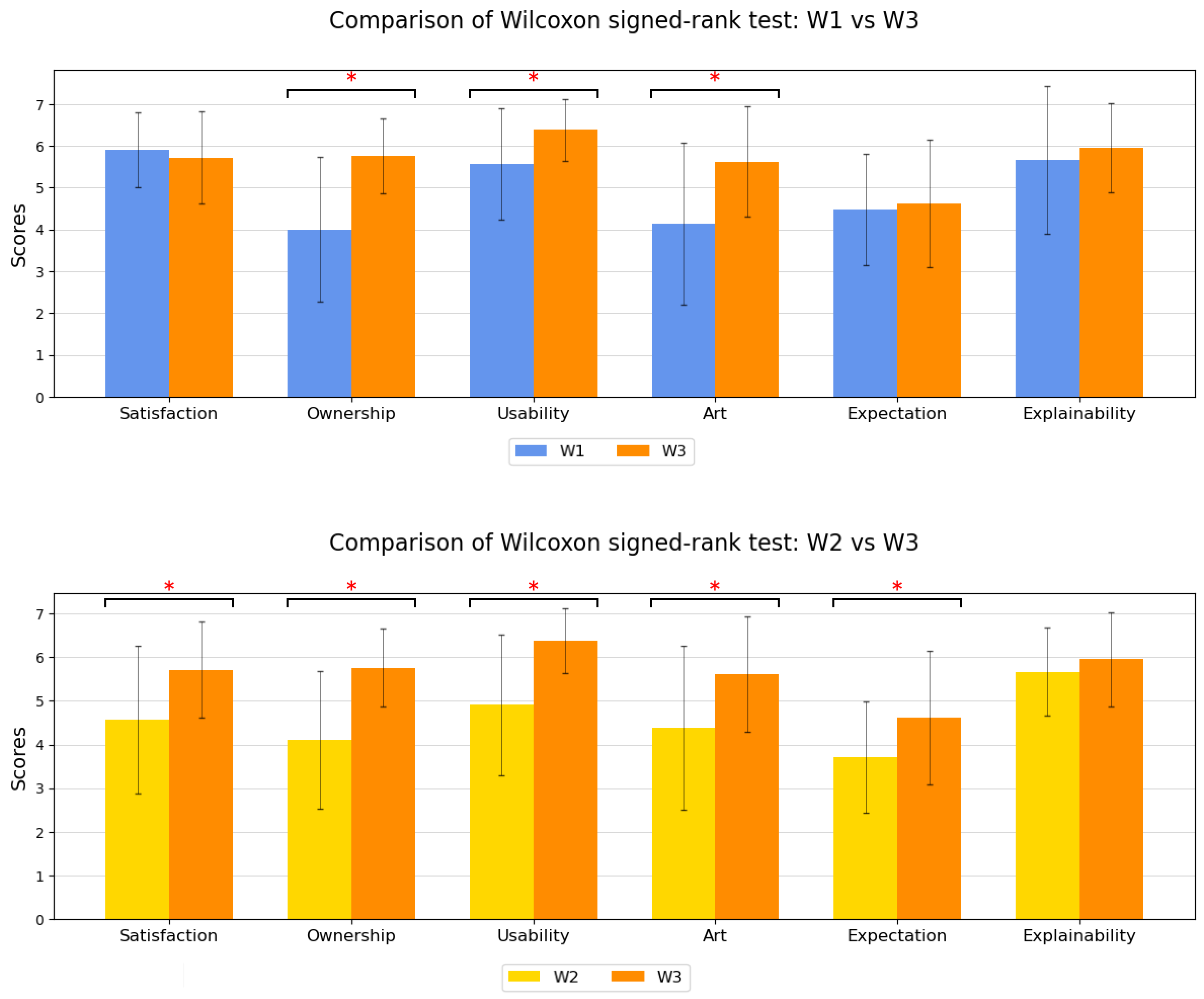}
        \caption{The charts compare Workflow 1 and Workflow 2 against Workflow 3 (LACE) using Likert scale scores across six categories. Red asterisks (*) indicate statistically significant differences between the workflows.}
        \label{fig:Likert-overall}
    \end{figure}

    \begin{figure}
        \centering
        \includegraphics[width=1.\linewidth]{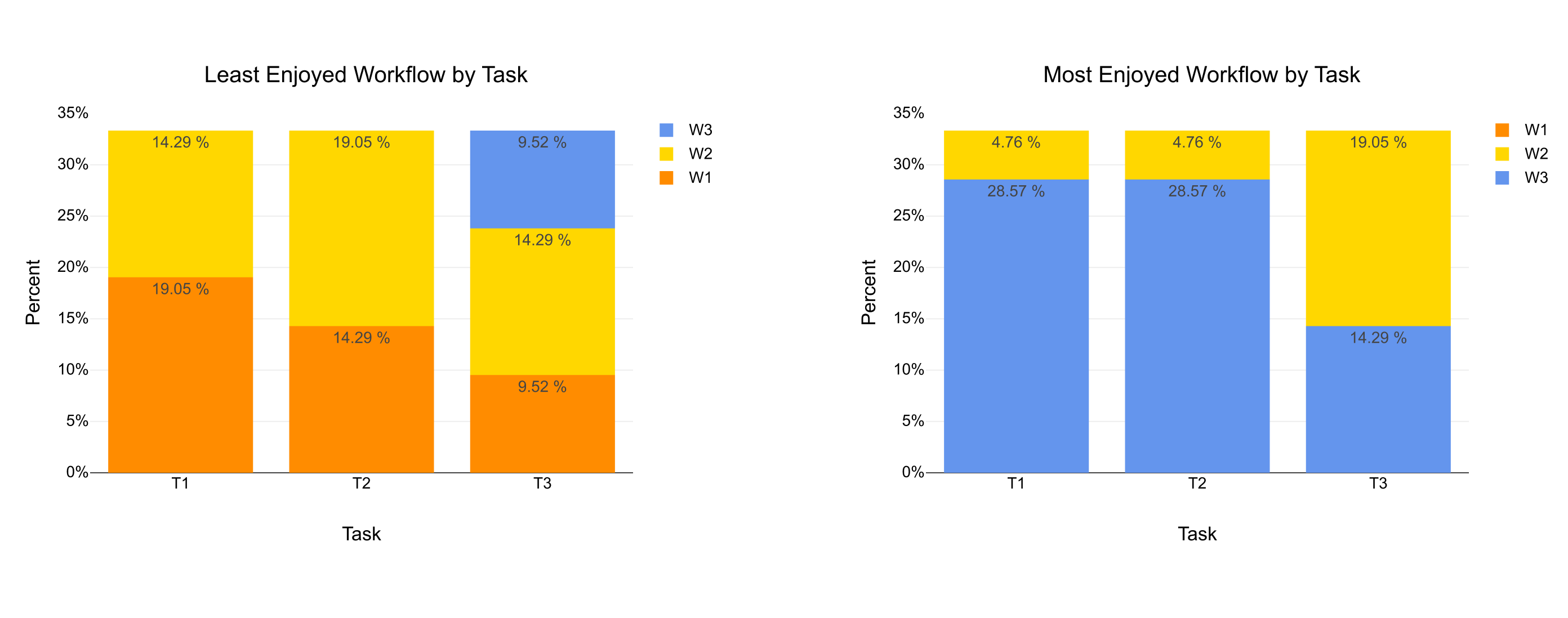}
        \caption{Preference of Workflow by Tasks}
        \label{fig:enjoy-w}
    \end{figure}

   The overall response to LACE highlights its advantages as a hybrid approach combining turn-taking and parallel workflows. Participants reported significant improvements in key metrics when using LACE (W3) compared to purely turn-taking workflows (W1 and W2). A Friedman test indicates significant differences between the workflows for satisfaction ($p$ = 0.039), ownership ($p$ = 0.009), usability ($p$ = 0.003), and art perception ($p$ = 0.005). Post-hoc pairwise comparisons using the Wilcoxon signed-rank test provide further insights. For ownership, LACE (W3) scores significantly higher than both W1 ( $z$ = -2.94 ,  $p$ = 0.002 ,  $r$ = 0.64 ) and W2 ( $z$ = -3.01 ,  $p$ = 0.001 ,  $r$ = 0.66 ), indicating greater user control and creative involvement. Similarly, usability scores for LACE are significantly higher compared to W1 ( $z$ = -2.54 ,  $p$ = 0.005 ,  $r$ = 0.56 ) and W2 ( $z$ = -3.33 ,  $p$ < 0.001 ,  $r$ = 0.73 ). For art perception, LACE is rated higher than both W1 ( $z$ = -3.24 ,  $p$ = 0.001 ,  $r$ = 0.71 ) and W2 ( $z$ = -2.37 ,  $p$ = 0.009 ,  $r$ = 0.52 ). These results suggest that integrating parallel and turn-taking modes in LACE enhances usability, ownership, and art perception compared to workflows relying solely on turn-taking (W1 and W2), as shown in Figure~\ref{fig:Likert-overall}.

    Participants rated Workflow 3 (LACE) significantly higher for usability (p < 0.005) and ownership (p < 0.001) compared to turn-taking workflows (W1 and W2). This reflects the added flexibility and creative control provided by LACE’s parallel/hybrid interaction design. Participants noted that W3’s ability to preserve layers and iterate independently enhanced their sense of ownership:
    \begin{quote}
        “I feel the most ownership over the last workflow, and the Photoshop layers allowed me to keep elements I liked from certain images” (P12).
    \end{quote}

    Workflow preferences varied by task type and the clarity of participants’ creative vision, shown in figure~\ref{fig:enjoy-w}. For tasks where participants lacked a clear creative vision, such as the more challenging T3 (design challenge), they relied on the AI to generate an initial output, making turn-taking workflows (W1 and W2) the preferred choice. These workflows provided structure and precision, enabling participants to iteratively refine the AI-generated results into coherent designs. Conversely, when participants had a clear creative vision, as in T1 (representational art) and T2 (abstract composition), they preferred LACE’s hybrid workflow (W3). Its parallel interaction mode facilitated real-time experimentation and creative freedom, allowing them to actively shape their work in alignment with their vision.
\section{LIMITATIONS}

Our study design grouped the parallel and hybrid (LACE) modes into a single workflow, making it difficult to discern how participants switched between turn-taking and parallel modes within LACE. Future studies should separate testing groups into distinct workflows for parallel, turn-taking, and hybrid modes to better understand user behavior and preferences for each interaction style.

Additionally, the insights gathered from our participants may not fully generalize to all users of systems like LACE. The participant pool primarily consisted of individuals with backgrounds in digital art or computer science, with varying levels of experience in AI tools and Photoshop. This variation likely influenced their ability to complete creative tasks and could affect the study’s outcomes. To improve the ecological validity of the findings, further research should include experienced professional artists who regularly work with advanced creative tools.

Finally, the study’s limitations include a small sample size, limited task time, and potential expertise mismatches between novices and advanced designers. Conducting in-situ testing with professional artists in real-world creative workflows would provide deeper insights into how systems like LACE support different creative processes and enhance their generalizability.
\section{DISCUSSION AND FUTURE WORK}

From our findings, we observed that the type of creative task and the stage of the creative process can influence the preferred participation mode in Human-AI collaboration. In the early stages of creation, when artists have less inspiration or need to generate initial ideas or compositions, turn-taking interaction appears more favorable. This is because turn-taking enables iterative refinement of text prompts and allows artists to provide high-level instructions to guide the AI. Conversely, in the later stages of creation, where the focus shifts to detail refinement and the scope of the work becomes more defined, real-time feedback and editing—facilitated by parallel interaction—may better support fine-grained control and precise adjustments.

However, these observations do not imply a clear division of specific modes for specific stages or tasks. The interplay between timing, creative tasks, and participation modes remains complex and context-dependent. Further investigation is needed to explore these dynamics in greater depth. Future work could aim to develop a comprehensive framework that better aligns participation modes with creative stages and task types, providing more actionable guidance for Human-AI co-creative systems.
\section{Conclusion}
Our findings highlight that turn-taking workflows are preferred during early-stage creation when users rely on AI for generating initial ideas, whereas parallel workflows support late-stage refinement by enabling fluid experimentation. Despite these insights, we observed variability in preferences across tasks and creative stages, underscoring the need for flexible systems like LACE that adapt to diverse creative workflows. This work contributes to understanding how interaction styles impact Human-AI collaboration and provides guidance for future systems supporting iterative and adaptive creative processes.

Our findings suggest that task type and user intent influence the preferred interaction mode in Human-AI collaboration. Turn-taking workflows were preferred for generating and refining initial ideas in tasks where participants lacked a clear creative vision. In contrast, parallel workflows, as supported by LACE, facilitated real-time experimentation and fine-grained control, making them more suitable for refinement when participants had a defined vision. Further studies are needed to validate these patterns and refine how participation modes align with creative stages and tasks.

\bibliographystyle{ACM-Reference-Format}
\bibliography{ref}


\begin{thebibliography}{18}


\ifx \showCODEN    \undefined \def \showCODEN     #1{\unskip}     \fi
\ifx \showDOI      \undefined \def \showDOI       #1{#1}\fi
\ifx \showISBNx    \undefined \def \showISBNx     #1{\unskip}     \fi
\ifx \showISBNxiii \undefined \def \showISBNxiii  #1{\unskip}     \fi
\ifx \showISSN     \undefined \def \showISSN      #1{\unskip}     \fi
\ifx \showLCCN     \undefined \def \showLCCN      #1{\unskip}     \fi
\ifx \shownote     \undefined \def \shownote      #1{#1}          \fi
\ifx \showarticletitle \undefined \def \showarticletitle #1{#1}   \fi
\ifx \showURL      \undefined \def \showURL       {\relax}        \fi
\providecommand\bibfield[2]{#2}
\providecommand\bibinfo[2]{#2}
\providecommand\natexlab[1]{#1}
\providecommand\showeprint[2][]{arXiv:#2}

\bibitem[Batley and Glithro(2024)]%
        {batley2024GFsynergy}
\bibfield{author}{\bibinfo{person}{Abigail Batley} {and} \bibinfo{person}{Richard Glithro}.} \bibinfo{year}{2024}\natexlab{}.
\newblock \showarticletitle{Exploring the Synergy of AI Generative Fill in Photoshop and the Creative Design Process Utilising Informal Learning}. In \bibinfo{booktitle}{\emph{DS 131: Proceedings of the International Conference on Engineering and Product Design Education (E\&PDE 2024)}}, \bibfield{editor}{\bibinfo{person}{Hilary Grierson}, \bibinfo{person}{Erik Bohemia}, {and} \bibinfo{person}{Lyndon Buck}} (Eds.). \bibinfo{publisher}{The Design Society}, \bibinfo{pages}{1--6}.
\newblock
\showISBNx{978-1-912254-200}
\showISSN{3005-4753}
\urldef\tempurl%
\url{https://doi.org/10.35199/EPDE.2024.1}
\showDOI{\tempurl}


\bibitem[Bozsó et~al\mbox{.}(2024)]%
        {semantic_segmentation_diffusion}
\bibfield{author}{\bibinfo{person}{Katica Bozsó}, \bibinfo{person}{András Béres}, {and} \bibinfo{person}{Bálint Gyires-Tóth}.} \bibinfo{year}{2024}\natexlab{}.
\newblock \showarticletitle{Semantic segmentation mask-guided diffusion models: A pathway to enriched datasets in autonomous systems}. \bibinfo{pages}{79--84}.
\newblock
\urldef\tempurl%
\url{https://doi.org/10.3311/WINS2024-014}
\showDOI{\tempurl}


\bibitem[Cao et~al\mbox{.}(2023)]%
        {cao2023animediffusion}
\bibfield{author}{\bibinfo{person}{Yu Cao}, \bibinfo{person}{Xiangqiao Meng}, \bibinfo{person}{PY Mok}, \bibinfo{person}{Xueting Liu}, \bibinfo{person}{Tong-Yee Lee}, {and} \bibinfo{person}{Ping Li}.} \bibinfo{year}{2023}\natexlab{}.
\newblock \showarticletitle{AnimeDiffusion: Anime Face Line Drawing Colorization via Diffusion Models}.
\newblock \bibinfo{journal}{\emph{arXiv preprint arXiv:2303.11137}} (\bibinfo{year}{2023}).
\newblock


\bibitem[Daniel(2022)]%
        {daniel2022creative}
\bibfield{author}{\bibinfo{person}{Ryan Daniel}.} \bibinfo{year}{2022}\natexlab{}.
\newblock \showarticletitle{The creative process explored: Artists’ views and reflections}.
\newblock \bibinfo{journal}{\emph{Creative Industries Journal}} \bibinfo{volume}{15}, \bibinfo{number}{1} (\bibinfo{year}{2022}), \bibinfo{pages}{3--16}.
\newblock


\bibitem[Fan et~al\mbox{.}(2019)]%
        {Fan2019Collabdraw}
\bibfield{author}{\bibinfo{person}{Judith~E. Fan}, \bibinfo{person}{Monica Dinculescu}, {and} \bibinfo{person}{David Ha}.} \bibinfo{year}{2019}\natexlab{}.
\newblock \showarticletitle{CollabDraw: An Environment for Collaborative Sketching with an Artificial Agent}. In \bibinfo{booktitle}{\emph{Proceedings of the 2019 Conference on Creativity and Cognition}}. \bibinfo{pages}{556--561}.
\newblock


\bibitem[Feng et~al\mbox{.}(2024)]%
        {feng2024item}
\bibfield{author}{\bibinfo{person}{Aosong Feng}, \bibinfo{person}{Weikang Qiu}, \bibinfo{person}{Jinbin Bai}, \bibinfo{person}{Kaicheng Zhou}, \bibinfo{person}{Zhen Dong}, \bibinfo{person}{Xiao Zhang}, \bibinfo{person}{Rex Ying}, {and} \bibinfo{person}{Leandros Tassiulas}.} \bibinfo{year}{2024}\natexlab{}.
\newblock \showarticletitle{An Item is Worth a Prompt: Versatile Image Editing with Disentangled Control}.
\newblock \bibinfo{journal}{\emph{arXiv preprint arXiv:2403.04880}} (\bibinfo{year}{2024}).
\newblock


\bibitem[Fischer(2004)]%
        {Fischer2004}
\bibfield{author}{\bibinfo{person}{Gerhard Fischer}.} \bibinfo{year}{2004}\natexlab{}.
\newblock \showarticletitle{Social creativity: turning barriers into opportunities for collaborative design}. In \bibinfo{booktitle}{\emph{Proceedings of the Eighth Conference on Participatory Design: Artful Integration: Interweaving Media, Materials and Practices - Volume 1}} (Toronto, Ontario, Canada) \emph{(\bibinfo{series}{PDC 04})}. \bibinfo{publisher}{Association for Computing Machinery}, \bibinfo{address}{New York, NY, USA}, \bibinfo{pages}{152–161}.
\newblock
\showISBNx{1581138512}
\urldef\tempurl%
\url{https://doi.org/10.1145/1011870.1011889}
\showDOI{\tempurl}


\bibitem[Gholami and Xiao(2024)]%
        {gholami2024streamliningimageeditinglayered}
\bibfield{author}{\bibinfo{person}{Peyman Gholami} {and} \bibinfo{person}{Robert Xiao}.} \bibinfo{year}{2024}\natexlab{}.
\newblock \bibinfo{title}{Streamlining Image Editing with Layered Diffusion Brushes}.
\newblock
\newblock
\showeprint[arxiv]{2405.00313}~[cs.CV]
\urldef\tempurl%
\url{https://arxiv.org/abs/2405.00313}
\showURL{%
\tempurl}


\bibitem[Ikegami and Iizuka(2007)]%
        {IkegamiIizuka2007}
\bibfield{author}{\bibinfo{person}{Takashi Ikegami} {and} \bibinfo{person}{Hiroyuki Iizuka}.} \bibinfo{year}{2007}\natexlab{}.
\newblock \showarticletitle{Turn-Taking Interaction as a Cooperative and Co-Creative Process}. In \bibinfo{booktitle}{\emph{Infant Behavior and Development}}. \bibinfo{pages}{278--288}.
\newblock


\bibitem[Lugmayr et~al\mbox{.}(2022)]%
        {lugmayr2022repaint}
\bibfield{author}{\bibinfo{person}{Andreas Lugmayr}, \bibinfo{person}{Martin Danelljan}, \bibinfo{person}{Andres Romero}, \bibinfo{person}{Fisher Yu}, \bibinfo{person}{Radu Timofte}, {and} \bibinfo{person}{Luc Van~Gool}.} \bibinfo{year}{2022}\natexlab{}.
\newblock \showarticletitle{Repaint: Inpainting using denoising diffusion probabilistic models}. In \bibinfo{booktitle}{\emph{Proceedings of the IEEE/CVF conference on computer vision and pattern recognition}}. \bibinfo{pages}{11461--11471}.
\newblock


\bibitem[{MidJourney}(2022)]%
        {MidJourney}
\bibfield{author}{\bibinfo{person}{{MidJourney}}.} \bibinfo{year}{2022}\natexlab{}.
\newblock \bibinfo{title}{MidJourney - Official Website}.
\newblock \bibinfo{howpublished}{\url{https://www.midjourney.com/}}.
\newblock
\newblock
\shownote{Accessed: 2024-09-15}.


\bibitem[Okada and Yokochi(2024)]%
        {okada2024process}
\bibfield{author}{\bibinfo{person}{Takeshi Okada} {and} \bibinfo{person}{Sawako Yokochi}.} \bibinfo{year}{2024}\natexlab{}.
\newblock \showarticletitle{Process Modification and Uncontrollability in an Expert Contemporary Artist's Creative Processes}.
\newblock \bibinfo{journal}{\emph{The Journal of Creative Behavior}} (\bibinfo{year}{2024}).
\newblock


\bibitem[Rezwana and Maher(2019)]%
        {Rezwana2019}
\bibfield{author}{\bibinfo{person}{Jeba Rezwana} {and} \bibinfo{person}{Mary~Lou Maher}.} \bibinfo{year}{2019}\natexlab{}.
\newblock \showarticletitle{A User-Centered Framework for Human-AI Co-Creativity}.
\newblock \bibinfo{journal}{\emph{Journal of Creative Computing}} \bibinfo{volume}{12}, \bibinfo{number}{3} (\bibinfo{year}{2019}), \bibinfo{pages}{45--60}.
\newblock


\bibitem[Rezwana and Maher(2023)]%
        {COFI}
\bibfield{author}{\bibinfo{person}{Jeba Rezwana} {and} \bibinfo{person}{Mary~Lou Maher}.} \bibinfo{year}{2023}\natexlab{}.
\newblock \showarticletitle{Designing Creative AI Partners with COFI: A Framework for Modeling Interaction in Human-AI Co-Creative Systems}. In \bibinfo{booktitle}{\emph{Proceedings of the CHI Conference on Human Factors in Computing Systems}}. \bibinfo{publisher}{ACM}, \bibinfo{pages}{1--28}.
\newblock


\bibitem[Rombach et~al\mbox{.}(2022)]%
        {StableDiffusion}
\bibfield{author}{\bibinfo{person}{Robin Rombach}, \bibinfo{person}{Andreas Blattmann}, \bibinfo{person}{Dominik Lorenz}, \bibinfo{person}{Patrick Esser}, {and} \bibinfo{person}{Bj\"orn Ommer}.} \bibinfo{year}{2022}\natexlab{}.
\newblock \showarticletitle{High-Resolution Image Synthesis With Latent Diffusion Models}. In \bibinfo{booktitle}{\emph{Proceedings of the IEEE/CVF Conference on Computer Vision and Pattern Recognition (CVPR)}}. \bibinfo{pages}{10684--10695}.
\newblock


\bibitem[Shneiderman(2002)]%
        {Shneiderman_2002}
\bibfield{author}{\bibinfo{person}{Ben Shneiderman}.} \bibinfo{year}{2002}\natexlab{}.
\newblock \showarticletitle{Creativity support tools}.
\newblock \bibinfo{journal}{\emph{Commun. ACM}} \bibinfo{volume}{45}, \bibinfo{number}{10} (\bibinfo{date}{Oct.} \bibinfo{year}{2002}), \bibinfo{pages}{116–120}.
\newblock
\showISSN{0001-0782}
\urldef\tempurl%
\url{https://doi.org/10.1145/570907.570945}
\showDOI{\tempurl}


\bibitem[van Berkel et~al\mbox{.}(2021)]%
        {Berkel_HAI}
\bibfield{author}{\bibinfo{person}{Niels van Berkel}, \bibinfo{person}{Mikael~B. Skov}, {and} \bibinfo{person}{Jesper Kjeldskov}.} \bibinfo{year}{2021}\natexlab{}.
\newblock \showarticletitle{Human-AI interaction: intermittent, continuous, and proactive}.
\newblock \bibinfo{journal}{\emph{Interactions}} \bibinfo{volume}{28}, \bibinfo{number}{6} (\bibinfo{date}{Nov.} \bibinfo{year}{2021}), \bibinfo{pages}{67–71}.
\newblock
\showISSN{1072-5520}
\urldef\tempurl%
\url{https://doi.org/10.1145/3486941}
\showDOI{\tempurl}


\bibitem[Yang et~al\mbox{.}(2024)]%
        {yang2024fine}
\bibfield{author}{\bibinfo{person}{Lingfeng Yang}, \bibinfo{person}{Yueze Wang}, \bibinfo{person}{Xiang Li}, \bibinfo{person}{Xinlong Wang}, {and} \bibinfo{person}{Jian Yang}.} \bibinfo{year}{2024}\natexlab{}.
\newblock \showarticletitle{Fine-grained visual prompting}.
\newblock \bibinfo{journal}{\emph{Advances in Neural Information Processing Systems}}  \bibinfo{volume}{36} (\bibinfo{year}{2024}).
\newblock


\end{thebibliography}
\balance
\clearpage
\appendix
\section{Appendix: Sampled Qualitative Results}
In this appendix, we present selected examples of participant-generated images from different workflows to provide qualitative insights into their effectiveness. Workflow 2 (W2) frequently resulted in lower-quality outputs due to participants' limited understanding of AI behaviors and its steep learning curve. Specifically, W2 relies on iterative sampling of latent vectors conditioned solely on text prompts from previous rounds, making coherent iteration difficult. Unlike Workflow 3 (W3), which allows users to visually edit decoded images directly in Photoshop, W2’s reliance on uninterpretable latent interactions can negatively affect user perceptions and creative outcomes. Despite these limitations, examining W2 remains valuable for understanding how indirect interactions influence users’ experiences.

Conversely, W3 consistently generated structured compositions aligned closely with participants' creative intentions. The clarity and directness of W3’s interaction—allowing parameter adjustments or direct canvas edits—significantly enhanced user agency. This direct manipulation makes user intentions explicit, facilitating easier debugging and refining of input conditions to achieve desired artistic outcomes. The following samples illustrate these differences clearly.

\begin{figure*}[p]
  \centering
  \includegraphics[width=.8\textwidth]{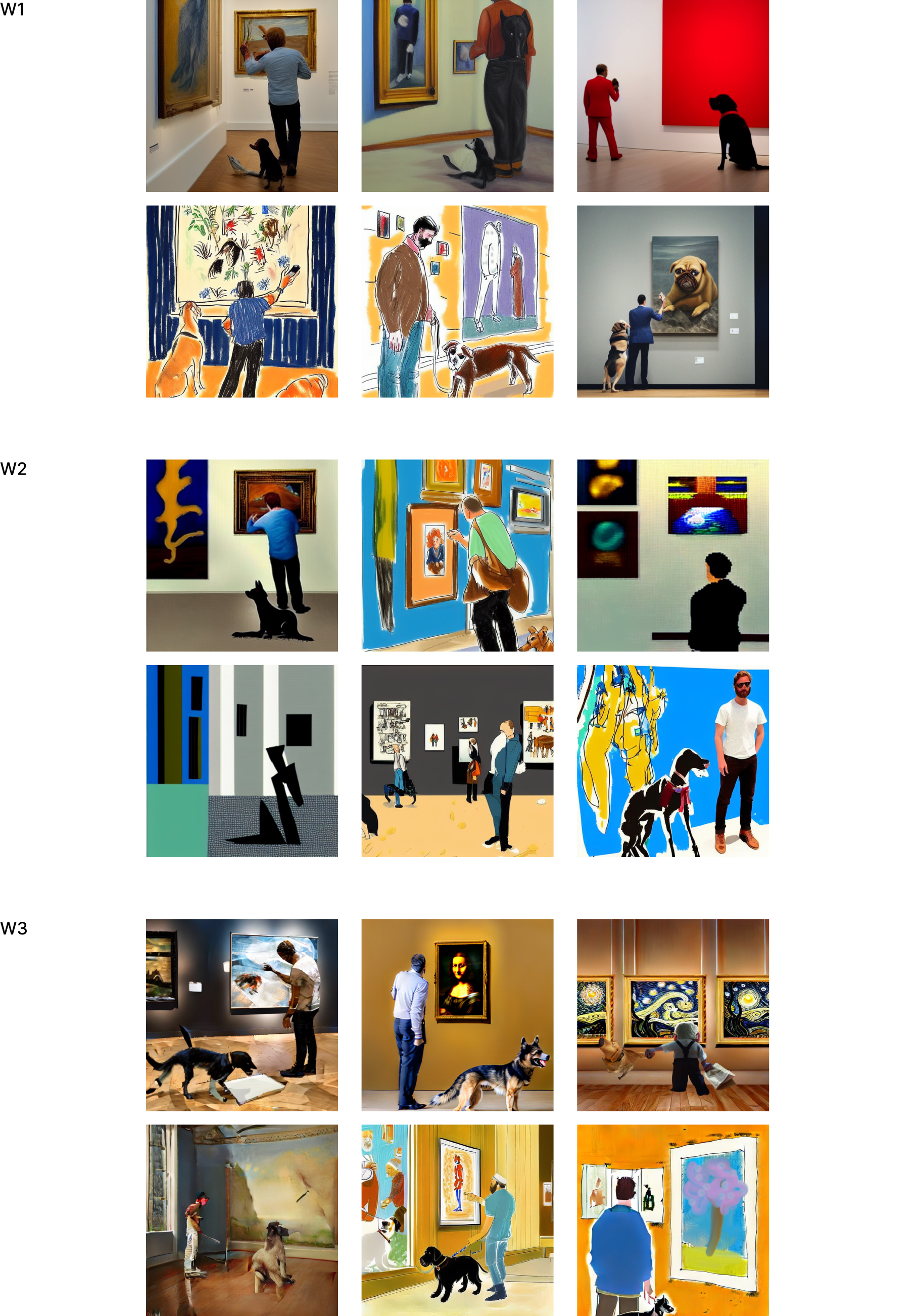}
  \caption{Results from Task 1. Prompt: \texttt{A man reaching for a painting in an art gallery, accompanied by a dog sniffing another artwork on the floor.''}}
  \label{fig:result_T1}
\end{figure*}

\begin{figure*}[p]
  \centering
  \includegraphics[width=0.8\textwidth]{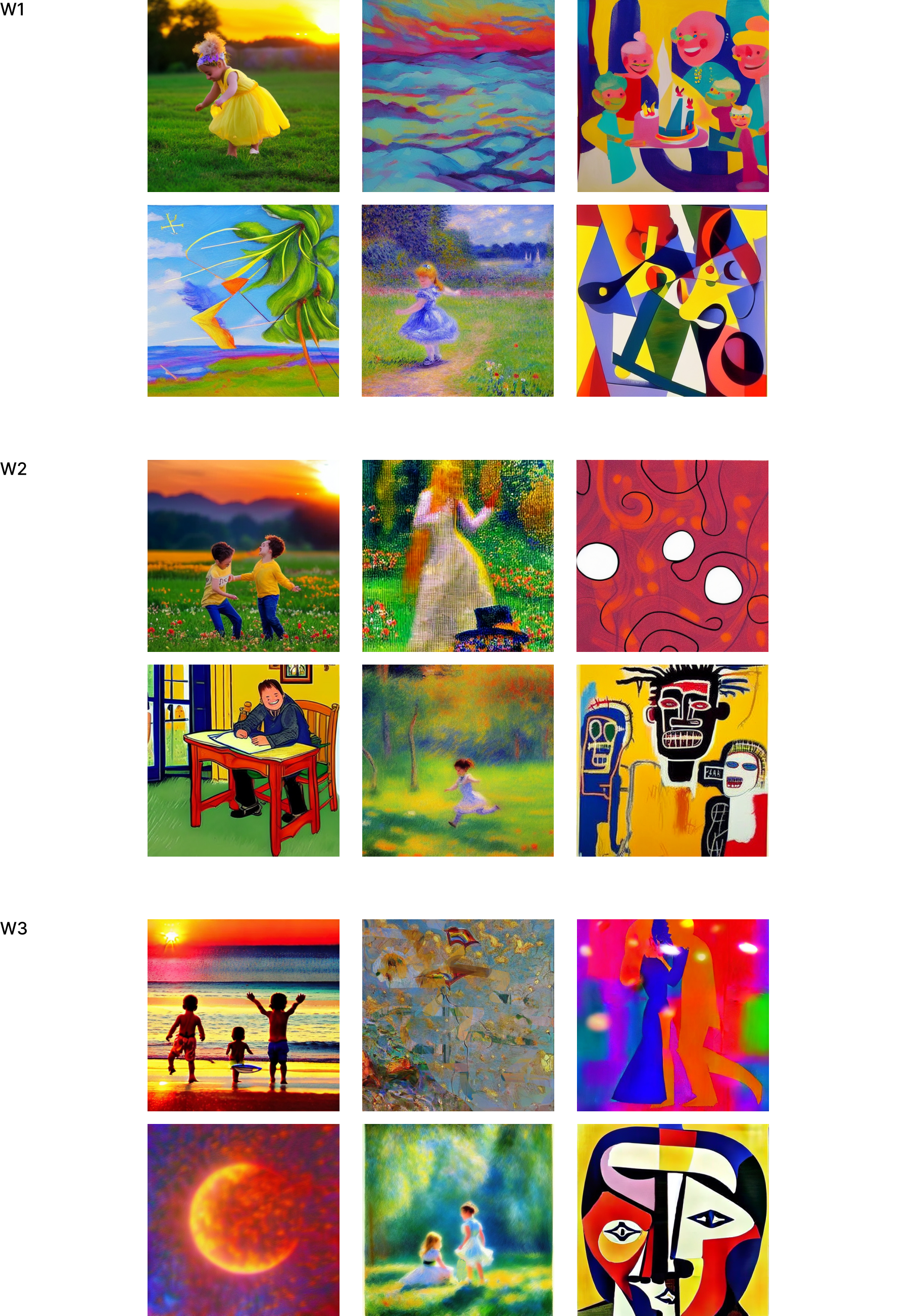}
  \caption{Results from Task 2. Prompt: \texttt{``An abstract composition that embodies the dynamics and motion associated with joy.''}}
  \label{fig:result_T2}
\end{figure*}

\begin{figure*}[p]
  \centering
  \includegraphics[width=0.8\textwidth]{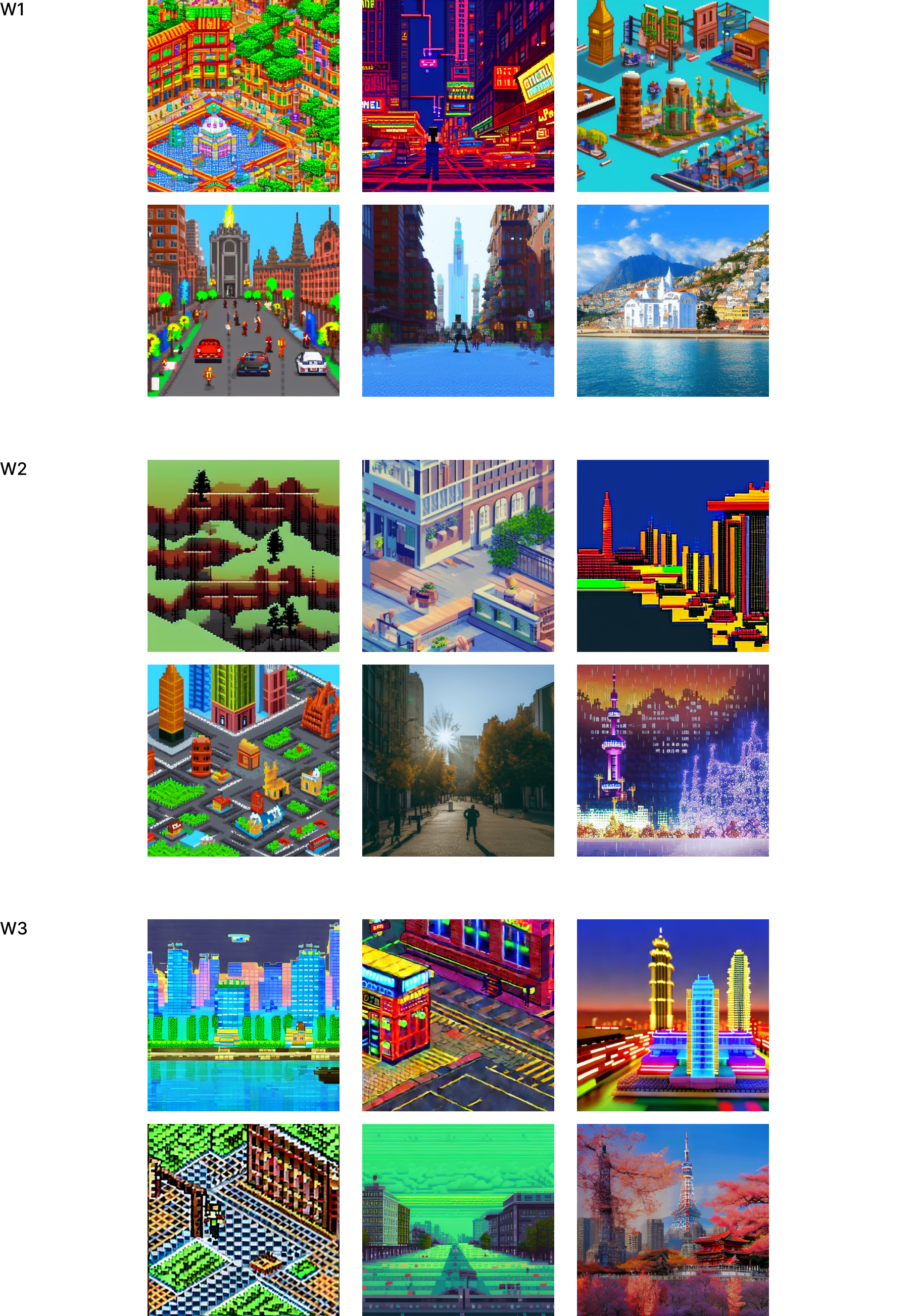}
  \caption{Results from Task 3. Prompt: \texttt{``A pixel art game scene with a bustling cityscape featuring assorted architectural styles.''}}
  \label{fig:result_T3}
\end{figure*}

\end{document}